\documentclass[11pt]{article}
\usepackage{coling08}
\usepackage{times}
\usepackage{latexsym}
\usepackage{color}
\usepackage{graphicx}
\usepackage{dcolumn}
\usepackage{bm}
\usepackage{epsfig}
\usepackage{amssymb,amsmath}
\usepackage{multirow}
\usepackage{rotating} 

\title{Scientific Paper Summarization Using Citation Summary Networks}

 \author{Vahed Qazvinian \\
   School of Information \\
   University of Michigan \\
   Ann Arbor, MI \\
   {\tt vahed@umich.edu} \And
   Dragomir R. Radev \\
 Department of EECS and\\
 School of Information\\
   University of Michigan\\
   Ann Arbor, MI\\
   {\tt radev@umich.edu}}

\date{}

\begin{document}
\maketitle

\vspace{-5mm}

\begin{abstract}
Quickly moving to a new area of research is painful for researchers
due to the vast amount of scientific literature in each field of
study. One possible way to overcome this problem is to summarize a
scientific topic. In this paper, we propose a model of summarizing a
single article, which can be further used to summarize an entire
topic. Our model is based on analyzing others' viewpoint of the target
article's contributions and the study of its citation summary network
using a clustering approach. 
\end{abstract}

\section{Introduction}

It is quite common for researchers to have to quickly move into a new area of research. For instance, someone trained in text generation may want to learn about parsing and someone  who knows summarization well, may need to learn about question answering.
In our work, we try to make this transition as painless as possible by automatically generating summaries of an entire research topic. This enables a researcher to find the chronological order and the progress in that particular field of study. An ideal such system will receive a topic of research, as the user query, and will return a summary of related work on that topic. In this paper, we take the first step toward building such a system.

Studies have shown that different citations to the same article often focus on different aspects of that article, while none alone may cover a full description of its entire contributions.
Hence, the set of citation summaries, can be a good resource to understand the main contributions of a paper and how that paper affects others.
The \emph{citation summary} of an article ($A$), as defined in \cite{Elkiss&al.07}, is a the set of citing sentences pointing to that article.
Thus, this source contains information about $A$ from others' point of view. Part of a sample citation summary is as follows:

\begin{description}
\begin{small}
\item {\it In the context of DPs, this \underline{edge based factorization} method was proposed by {\bf (Eisner, 1996)}.} \vspace{-2mm}
\item  {\it {\bf Eisner (1996)} gave a  \underline{generative model} with a  \underline{cubic parsing} algorithm based on an edge factorization of trees.}\vspace{-2mm}
\item  {\it Eisner {\bf (Eisner, 1996)} proposed an \underline{$O(n^3)$ parsing algorithm for PDG}.}\vspace{-2mm}
\item  {\it If the parse has to be projective, Eisner's \underline{bottom-up-span algorithm} {\bf (Eisner, 1996)} can be used for the search.} \vspace{-2mm}
\end{small}
\end{description}
The problem of summarizing a whole scientific topic, in its simpler form, may reduce to summarizing one particular article. A citation summary can be a good resource to make a summary of a target paper. Then using each paper's summary and some knowledge of the citation network, we'll be able to generate a summary of an entire topic. Analyzing citation networks is an important component of this goal, and has been widely studied before \cite{Newman01a}.

Our main contribution in this paper is to use citation summaries and network analysis techniques to produce a summary of a single scientific article as a framework for future research on topic summarization.
Given that the citation summary of any article usually has more than a few sentences, the main challenge of this task is to find a \emph{subset} of these sentences that will lead to a better and shorter summary.

\vspace{-3mm}
\subsection{Related Work}\label{sect:relatedwork}

Although there has been work on analyzing citation and collaboration networks \cite{teufel2006,Newman01a} and scientific article summarization \cite{simone2002}, to the knowledge of the author there is no previous work that study the text of the citation summaries to produce a summary.
\cite{Bradshaw03,Bradshaw02} get benefit from citations to determine the content of articles and introduce ``Reference Directed Indexing'' to improve the results of a search engine.

In other work, \cite{NanbaKO00,NanbaEtal04} analyze citation sentences and automatically categorize citations into three groups using 160 pre-defined
phrase-based rules. This categorization is then used to build a tool for survey generation.
\cite{Nanba&Okumura99} also discuss the same citation categorization to support a system for writing a survey.
\cite{Nanba&Okumura99,NanbaKO00} report that co-citation implies similarity by showing that the textual similarity of co-cited papers is
proportional to the proximity of their citations in the citing article.

Previous work has shown the importance of the citation summaries in understanding what a paper says.
The \emph{citation summary} of an article $A$ is the set of sentences in other articles which cite $A$.
\cite{Elkiss&al.07} performed a large-scale study on citation summaries and their importance.
They conducted several experiments on a set of $2,497$ articles from the free PubMed Central (PMC) repository\footnote{ http://www.pubmedcentral.gov}.
Results from this experiment confirmed that the ``Self Cohesion'' \cite{Elkiss&al.07} of a citation summary of an article is consistently higher
than the that of its abstract.
\begin{table}
\begin{center}
\begin{scriptsize}
\begin{tabular}{|l|l c c r|}
\hline
 Cluster & Nodes  &  &  & Edges\\
\hline
DP & 167  &  &  & 323 \\
PBMT & 186 & & &  516  \\
Summ & 839 & &  & 1425  \\
QA & 238 & &  &  202   \\
TE & 56 & &  & 44  \\
\hline
\end{tabular}
\end{scriptsize}
\caption{Clusters and their citation network size}\label{tbl:cluster}
\end{center}
\end{table}
\cite{Elkiss&al.07} also conclude that citation summaries are more focused than abstracts, and that they contain additional information that does not appear in abstracts. \cite{Kupiec95} use the abstracts of scientific articles as a target summary, where they use 188 Engineering Information summaries that are mostly indicative in nature. Abstracts tend to summarize the document’s topics well, however, they don't include much use of metadata. \cite{kan2002} use annotated bibliographies to cover certain aspects of summarization and suggest guidelines that summaries should also include metadata and critical document features as well as the prominent content-based features.

Siddharthan and Teufel describe a new task to decide the scientific attribution of an article \cite{simone2007} and show high human agreement as well as an improvement in the performance of Argumentative Zoning \cite{teufel2005}. Argumentative Zoning is a rhetorical classification task, in which sentences are labeled as one of {Own, Other, Background, Textual, Aim, Basis, Contrast} according to their role in the author's argument. These all show the importance of citation summaries and the vast area for new work to analyze them to produce a  summary for a given topic.

\begin{table*}
\begin{center}
{\scriptsize
\begin{tabular}{|l|l|l|c|c|}
\hline
 &  ACL-ID & Title & Year & CS Size\\
\hline
\multirow{5}{*}{\rotatebox{90}{\mbox{DP}}} &  C96-1058 & \small Three New Probabilistic Models For Dependency Parsing: An Exploration \normalsize&  1996 &  66\\
   &  P97-1003 & \small Three Generative, Lexicalized Models For Statistical Parsing \normalsize& 1997 & 55\\
   &  P99-1065 & \small A Statistical Parser For Czech \normalsize& 1999 & 54 \\
   &  P05-1013 & \small Pseudo-Projective Dependency Parsing \normalsize& 2005 & 40\\
   &  P05-1012 & \small On-line Large-Margin Training Of Dependency Parsers \normalsize& 2005 & 71\\
\hline
\multirow{5}{*}{\rotatebox{90}{\mbox{PBMT}}} & N03-1017 &  \small Statistical Phrase-Based Translation \normalsize & 2003 & 180\\
     & W03-0301 & \small An Evaluation Exercise For Word Alignment \normalsize & 2003 & 14 \\
     & J04-4002 & \small The Alignment Template Approach To Statistical Machine Translation\normalsize & 2004 & 50\\
     & N04-1033 & \small Improvements In Phrase-Based Statistical Machine Translation \normalsize &  2004 &  24\\
     & P05-1033 & \small A Hierarchical Phrase-Based Model For Statistical Machine Translation \normalsize & 2005 & 65\\
\hline	 	
\multirow{5}{*}{\rotatebox{90}{\mbox{Summ}}} &  A00-1043  & \small Sentence Reduction For Automatic Text Summarization \normalsize & 2000 & 19 \\
      & A00-2024 & \small Cut And Paste Based Text Summarization \normalsize & 2000 & 20   \\
      & C00-1072 & \small The Automated Acquisition Of Topic Signatures For Text Summarization \normalsize & 2000 & 19   \\
      & W00-0403 &  \small Centroid-Based Summarization Of Multiple Documents: Sentence Extraction, ...\normalsize & 2000& 31 \\
      & W03-0510 &  \small The Potential And Limitations Of Automatic Sentence Extraction For Summarization \normalsize & 2003 & 15   \\
\hline	
\multirow{5}{*}{\rotatebox{90}{\mbox{QA}}} &  A00-1023 & \small A Question Answering System Supported By Information Extraction \normalsize & 2000 & 13 \\
      & W00-0603 & \small A Rule-Based Question Answering System For Reading Comprehension Tests \normalsize & 2002 & 19 \\
      & P02-1006  & \small Learning Surface Text Patterns For A Question Answering System \normalsize & 2002 & 74 \\
      & D03-1017  & \small Towards Answering Opinion Questions: Separating Facts From Opinions ... \normalsize & 2003 & 42\\
      & P03-1001  & \small Offline Strategies For Online Question Answering: Answering Questions Before They Are Asked \normalsize & 2003 & 27\\

\hline
\multirow{5}{*}{\rotatebox{90}{\mbox{TE}}}   &  D04-9907 & \small Scaling Web-Based Acquisition Of Entailment Relations \normalsize & 2004 & 12 \\
      & H05-1047 & \small A Semantic Approach To Recognizing Textual Entailment \normalsize & 2005 & 8 \\
      &  H05-1079  & \small Recognising Textual Entailment With Logical Inference \normalsize & 2005 & 9\\
      &  W05-1203 & \small Measuring The Semantic Similarity Of Texts \normalsize & 2005 & 17 \\
      &  P05-1014 & \small The Distributional Inclusion Hypotheses And Lexical Entailment \normalsize & 2005 & 10 \\
\hline
\end{tabular}
\caption{Papers chosen from clusters for evaluation, with their publication year, and citation summary size}\label{tbl:paperchoice}
}
\end{center}
\end{table*}
\vspace{-2mm}
\section{Data}\label{sect:Data}

The ACL Anthology is a collection of papers from the Computational Linguistics journal, and proceedings from ACL conferences and workshops and includes almost $11,000$ papers.
To produce the {\bf A}CL {\bf A}nthology {\bf N}etwork (AAN),
\cite{Joseph&Radev07} manually performed some preprocessing tasks
including parsing references and building the network metadata, the
citation, and the author collaboration networks. 

The full AAN includes all citation and collaboration data within the ACL papers, with the citation network consisting of $8,898$ nodes and $38,765$ directed edges.

\vspace{-3mm}
\subsection{Clusters}

We built our corpus by extracting small clusters from the AAN data. Each cluster includes papers with a specific phrase in the title or content.
We used a very simple approach to collect papers of a cluster,  which are likely to be talking about the same topic.
Each cluster consists of a set of articles, in which the topic phrase is matched within the title or the content of papers in AAN. In particular, the five clusters that we collected this way, are: {\bf D}ependency {\bf P}arsing (DP), {\bf P}hrased {\bf B}ased {\bf M}achine {\bf T}ranslation (PBMT),  Text {\bf Summ}arization (Summ), {\bf Q}uestion {\bf A}nswering (QA), and  {\bf T}extual {\bf E}ntailment (TE).
Table \ref{tbl:cluster} shows the number of articles and citations in each cluster.
For the evaluation purpose we chose five articles from each cluster.
Table \ref{tbl:paperchoice} shows the title, year, and citation summary size for the 5 papers chosen from each cluster. The citation summary size of a paper is the size of the set of citation sentences that cite that paper.

\vspace{-3mm}
\section{Analysis}

\subsection{Fact Distribution}\label{sect:factdist}

We started with an annotation task on 25 papers, listed in Table \ref{tbl:paperchoice}, and asked a number of annotators to read the citation summary of each paper and extract a list of the main contributions of that paper.  Each item on the list is a  \emph{non-overlapping contribution} ({\bf \emph{fact}}) perceived by reading the citation summary.
We asked the annotators to focus on the citation summary to do the task and not on their background on this topic.

As our next step we manually created the union of the shared and similar facts by different annotators to make a list of facts for each paper. This fact list made it possible to review all sentences in the citation summary to see which facts each sentence contained. There were also some unshared facts, facts that only appear in one annotator's result, which we ignored for this paper.

Table \ref{tbl:collins99}  shows the shared and  unshared facts
extracted by four annotators for P99-1065.

 \begin{table}
 \begin{center}
{\small
 \begin{tabular}{|l | l  r|}
\hline
& Fact & Occurrences\\\hline
\multirow{7}{*}{\rotatebox{90}{\mbox{Shared}}}& $f_4$: ``Czech DP''        &        10\\
& $f_1$: ``lexical rules''& 6\\
& $f_3$: ``POS/ tag classification''  & 6\\
& $f_2$: ``constituency parsing'' & 5\\
& $f_5$: ``Punctuation''        &     2\\
& $f_6$: ``Reordering Technique''   & 2\\
& $f_7$: ``Flat Rules''        &      2\\
 \hline
 \multirow{7}{*}{\rotatebox{90}{\mbox{Unshared}}}	& ``Dependency conversion'' &	 \\
         & ``80\% UAS'' & \\
 	&  ``97.0\% F-measure''& \\
 	& ``Generative model'' 	& \\
 	& `Relabel coordinated phrases''& \\
 	& `Projective trees''	 & \\
 	& ``Markovization''	& \\
\hline
 \end{tabular}
}
 \caption{Facts of P99-1065}\vspace{-2mm}\label{tbl:collins99}
 \end{center}
 \end{table}

The manual annotation of P99-1065 shows that the fact ``Czech DP''
appears in 10 sentences out of all 54 sentences in the citation
summary. This shows the importance of this fact, and that ``Dependency
Parsing of Czech'' is one of the main contributions of this
paper. Table \ref{tbl:collins99} also shows the number of times each
shared fact appears in P99-1065's citation summary sorted by
occurrence. 

After scanning through all sentences in the citation summary, we can
come up with a {\bf \emph{fact distribution matrix}} for an
article. The rows of this matrix represent sentences in the citation
summary and the columns show facts. A $1$ value in this matrix means
that the sentence covers the fact. The matrix $D$ shows the fact
distribution of P99-1065. IDs in each row show the citing article's
ACL ID, and the sentence number in the citation summary. These
matrices, created using annotations, are particularly useful in the
evaluation process.


\scriptsize
\begin{displaymath}
D= \left(
\begin{array}{c|ccccccc}
& f_1 & f_2 & f_3& f_4 & f_5& f_6 & f_7\\ \hline
\tiny W06\textrm{-}2935\textrm{:}1      &1       &0       &0       &0       &0       &0       &0\\
W06\textrm{-}2935\textrm{:}2      &0  	   &0       &0       &0       &0       &0       &0\\
W06\textrm{-}2935\textrm{:}3  	  &0       &0       &1       &1       &0       &0       &0\\
W06\textrm{-}2935\textrm{:}4      &0       &0       &0       &0       &0       &0       &1\\
W06\textrm{-}2935\textrm{:}5      &0       &0       &0       &0       &0       &0       &0\\
W06\textrm{-}2935\textrm{:}6      &0       &0       &0       &0       &1       &0       &0\\
W05\textrm{-}1505\textrm{:}7      &0       &1       &0       &1       &0       &0       &0\\
W05\textrm{-}1505\textrm{:}8      &0       &0       &0       &0       &0       &1       &0\\
\vdots  & \vdots & & &\vdots & & &\vdots \\
W05\textrm{-}1518\textrm{:}54     &0       &0       &0       &0       &0       &0       &0\\
\end{array}
\right)
\end{displaymath}
\normalsize


\subsection{Similarity Measures}\label{sect:similarity}

We want to build a network with citing sentences as nodes and similarities of two sentences as edge weights.
We'd like this network to have a nice community structure, whereby each cluster corresponds to a fact.
So, a similarity measure in which we are interested is the one which results in high values for pairs of sentences that cover the same facts. On the other hand, it should return a low value for pairs that do not share a common contribution of the target article. 

The following shows two sample sentences from P99-1065 that cover the same fact and we want the chosen similarity measure to return a high value for them:
\begin{description}
\begin{small}
\item {\it So, Collins et al (1999) proposed a \underline{tag classification} for parsing the \underline{Czech} treebank.}  \vspace{-2mm}
\item {\it The \underline{Czech parser} of Collins et al (1999) was run on a different data set... . }\vspace{-2mm}
\end{small}
\end{description}
Conversely, we'd like the similarity of the two following sentences that cover no shared facts, to be quite low:
\begin{description}
\begin{small}
\item {\it Collins (1999) explicitly added features to his parser to improve \underline{punctuation dependency parsing} accuracy.}\vspace{-2mm}
\item {\it The trees are then transformed into Penn Treebank style constituencies-  using the technique described in (Collins et al, 1999). }\vspace{-2mm}
\end{small}
\end{description}
We used P99-1065 as the training sample, on which similarity metrics were trained, and left the others for the test. To evaluate a similarity measure for our purpose we use a simple approach. For each measure, we sorted the similarity values of all pairs of sentences in P99-1065's citation summary in a descending order. Then we simply counted the number of pairs that cover the same fact (out of $172$ such fact sharing pairs) in the top $100$, $200$ and $300$ highly similar ones out of total $2,862$ pairs.  Table \ref{tbl:similarities} shows the number of fact sharing pairs among the top highest similar pairs.
 \begin{table}
 \begin{center}
{\scriptsize
 \begin{tabular}{|l|c|c|c|}
\hline
Measure & Top $100$ &  Top $200$ &  Top $300$ \\
\hline
tf-idf (General) & $34$  & $66$ & $74$\\
tf-idf (AAN) & $34$ & $56$ & $74$  \\
LCSS & $26$ & $37$ & $54$\\
tf & $24$ & $34$ & $46$ \\
tf2gen  & $13$ &$26$ & $35$\\
tf-idf (DP) & $16$ &$26$ & $28$ \\
Levenshtein & $2$ & $9$ & $16$ \\
\hline
\end{tabular}
\caption{Different similarity measures and their performances in favoring fact-sharing sentences. Each column shows the number of fact-sharing pairs among top highly similar pairs.}\label{tbl:similarities}
}
\end{center}
\end{table}
Table \ref{tbl:similarities} shows how cosine  similarity that uses a tf-idf measure outperforms the others. We tried three different policies for computing IDF values to compute cosine similarity: a general IDF, an AAN-specific IDF where IDF values are calculated only using the documents of AAN, and finally DP-specific IDF in which we only used all-DP data set. Table \ref{tbl:similarities} also shows the results for an asymmetric similarity measure, generation probability \cite{erkan:2006:HLT-NAACL06-Main} as well as two string edit distances: the longest common substring and the Levenshtein distance \cite{levenshtein66}.
\vspace{-2mm}
\section{Methodology}

In this section we discuss our graph clustering method for article summarization, as well as other baseline methods used for comparisons.
\subsection{Network-Based Clustering}

\begin{figure}
\centering
\includegraphics[scale=0.1315]{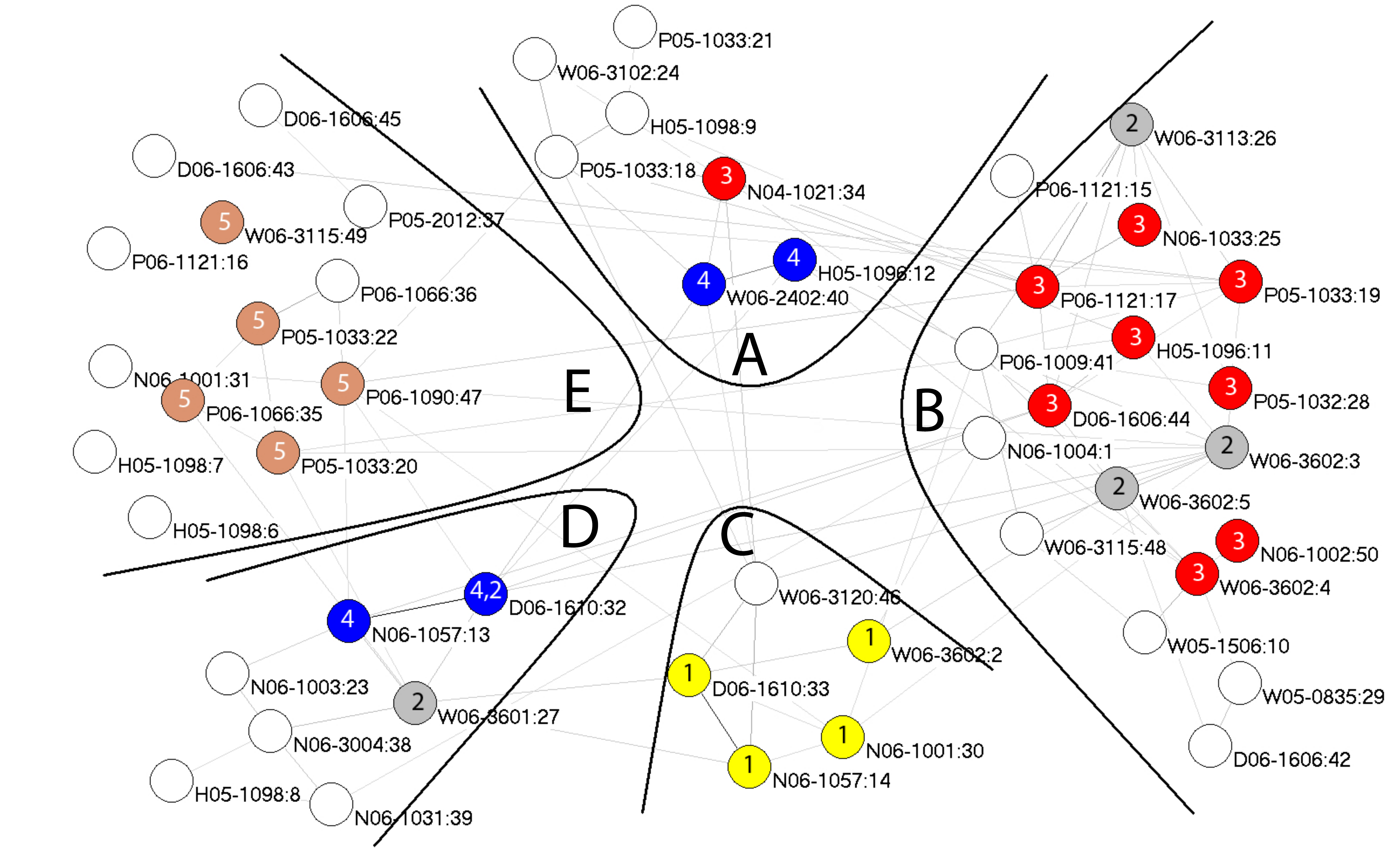}
\caption{Each node is a sentence in the citation summary for paper J04-4002. Colors (numbers) represent facts and boundaries show the clustering result}\label{fig:final1}
\end{figure}

\emph{The Citation Summary Network} of an article $A$ is a network in which each sentence in the citation summary of $A$ is a node. This network is a \emph{complete undirected weighted graph} where the weight of an edge between two nodes shows the \emph{similarity} of the two corresponding sentences of those nodes.
The similarity that we use, as described in section \ref{sect:similarity}, is such that sentences with the same facts have high similarity values. In other words, strong edges in the citation summary network are likely to indicate a shared fact between two sentences.\\
A graph clustering method tries to cluster the nodes of a graph in a way that the average intra-cluster similarity is maximum and the average inter-cluster similarity is minimum. To find the communities in the citation summary network we use \cite{Clauset04}, a  hierarchical agglomeration algorithm which works by greedily optimizing the modularity in a linear running time for sparse graphs.\\
To evaluate how well the clustering method works, we calculated the \emph{purity} for the clusters found of each paper. Purity \cite{manning07} is a method in which each cluster is assigned to the class with the majority vote in the cluster, and then the accuracy of this assignment is measured by dividing the number of correctly assigned documents by $N$. More formally:
\vspace{-2mm}
\begin{displaymath} \mbox{purity}( \Omega,\mathbb{C} ) = \frac{1}{N} \sum_k \max_j \vert\omega_k \cap c_j\vert \end{displaymath}
where $\Omega = \{ \omega_1, \omega_2, \ldots, \omega_K \}$ is the set of clusters and $\mathbb{C} = \{ c_1,c_2,\ldots,c_J \}$ is the set of classes. $\omega_k$ is interpreted as the set of documents in  $\omega_k$ and $c_j$ as the set of documents in $c_j$. For each evaluated article, Table  \ref{tbl:purity} shows the number of real facts ($|\mathbb{C}|= J$), the number of clusters ($|\Omega| = K$) and $purity( \Omega,\mathbb{C} )$ for each evaluated article. Figure \ref{fig:final1} shows the clustering result for J04-4002, in which each color (number) shows a real fact, while the boundaries and capital labels show the clustering result.

\begin{table}
\begin{center}
{\scriptsize
\begin{tabular}{|l|l|c|c|c|}
\hline
 &  ACL-ID & \#Facts $|\mathbb{C}|$ & \#Clusters $|\Omega|$ &  $Purity( \Omega,\mathbb{C} )$\\
\hline
\multirow{5}{*}{\rotatebox{90}{\mbox{DP}}} &  C96-1058 &  $4$ & $4$  & $0.636$ \\
   &  P97-1003 & $5$	&$5$	&$0.750$\\
   &  P99-1065 & $7$	&$7$	&$0.724$ \\
   &  P05-1013 & $5$   &$3$	&$0.689$ \\
   &  P05-1012 & $7$	&$5$	&$0.500$ \\
\hline
\multirow{5}{*}{\rotatebox{90}{\mbox{PBMT}}} & N03-1017 & $8$  &$4$  &$0.464$ \\
     & W03-0301 & $3$	&$3$	&$0.777$ \\
     & J04-4002 & $5$	&$5$	&$0.807$ \\
     & N04-1033 & $5$	&$4$	&$0.615$ \\
     & P05-1033 & $6$	&$5$	&$0.650$ \\
\hline	
\multirow{5}{*}{\rotatebox{90}{\mbox{Summ}}} & A00-1043  & $5$  & $4$ & $0.812$\\
      & A00-2024  &  $5$ & $2$  & $0.333$\\
      & C00-1072  &  $3$ & $4$  & $0.857$\\
      & W00-0403 &   $6$ & $4$  & $0.682$\\
      & W03-0510 &  $4$ & $3$ & $0.727$\\
\hline	
\multirow{5}{*}{\rotatebox{90}{\mbox{QA}}} & A00-1023 & $3$  & $2$ & $0.833$\\
	& W00-0603  & $7$  & $4$ & $0.692$\\
	& P02-1006  & $7$  & $5$ & $0.590$\\
	& D03-1017  & $7$  & $4$ & $0.500$ \\
        & P03-1001  & $6$  & $4$ & $0.500$\\

\hline	
\multirow{5}{*}{\rotatebox{90}{\mbox{TE}}}  &  D04-9907  & $7$  & $3$ & $0.545$\\
      & H05-1047 & $4$  & $3$ & $0.833$ \\
      &  H05-1079  & $5$  & $3$ & $0.625$\\
      &  W05-1203  & $3$  & $3$ & $0.583$\\
      &  P05-1014  & $4$  & $2$ & $0.667$\\
\hline		
\end{tabular}
\caption{Number of real facts, clusters and purity for each evaluated article}\label{tbl:purity}
}
\end{center}
\end{table}
\normalsize
 \begin{table*}
 \begin{center}
{\scriptsize
 \begin{tabular}{|@{}l@{} | l|}\hline
 ID & Sentence \\ \hline
\multicolumn{2}{|c|}{C-RR}\\ \hline
W05-1505:9 &	3 Constituency Parsing for Dependency Trees A pragmatic justification for using constituency- based parser in order\\
& to predict dependency struc- tures is that currently the best Czech dependency- tree parser  is a constituency-based  parser (Collins et al, 1999; Zeman, 2004). \\

W04-2407:27 &	However, since most previous studies instead use the mean attachment score per word (Eisner, 1996; Collins et al, 1999), we will give this measure as well. \\

J03-4003:33 &	3 We find lexical heads in Penn Treebank data using the rules described in Appendix A of Collins (1999). \\

H05-1066:51 &	Furthermore, we can also see that the MST parsers perform favorably compared to the more powerful \\
& lexicalized phrase-structure parsers, such as those presented by Collins et al  (1999) and Zeman (2004) that use expensive O(n5) parsing al- gorithms.\\

E06-1011:21 &	5.2 Czech Results For the Czech data, we used the predefined train- ing, development and testing split\\
& of the Prague Dependency Treebank (Hajiˇc et al, 2001), and the automatically generated POS tags supplied with the data,\\ &which we reduce to the POS tag set from Collins et al (1999). \\

\hline \hline
\multicolumn{2}{|c|}{C-Lexrank}\\ \hline
P05-1012:16 &	The Czech parser of Collins et al (1999) was run on a different data set and most other dependency parsers are evaluated using English. \\

W04-2407:26 &	More precisely, parsing accuracy is measured by the attachment score, which is \\
& a standard measure used in studies of dependency parsing (Eisner, 1996; Collins et al, 1999). \\

W05-1505:14 &	In an attempt to extend a constituency-based pars- ing model to train on dependency trees,\\
&  Collins transforms the PDT dependency trees into con- stituency trees (Collins et al, 1999). \\

P06-1033:31 &	More specifi- cally for PDT, Collins et al (1999) relabel coordi- nated phrases after  converting dependency struc- tures to phrase \\
& structures, and Zeman (2004) uses a kind of pattern matching, based on frequencies of the parts-of-speech of conjuncts and conjunc- tions. \\

P05-1012:17 &	In par- ticular, we used the method of Collins et al (1999) to simplify part-of-speech tags since\\
& the rich tags used by Czech would have led to a large but rarely seen set of POS features. \\
\hline
 \end{tabular}
 \caption{System Summaries for P99-1065. (a) Using C-RR, (b) using C-Lexrank with length of 5 sentences}\label{tbl:systemSumm}
}
 \end{center}
 \end{table*}

\vspace{-2mm} 
\subsubsection{Sentence Extraction}\label{sect:extraction}
\vspace{-2mm}
Once the graph is clustered and communities are formed, to build a summary we extract sentences from the clusters. We tried these two different simple methods:
\begin{itemize}
 
 \item {\bf {\it Cluster Round-Robin (C-RR)}}\\ We start with the largest cluster, and extract sentences in the order they appear in each cluster. So we extract first, the first sentences from each cluster, then the second ones, and so on, until we reach the summary length limit $|S|$.
Previously, we mentioned that facts with higher weights appear in greater number of sentences, and clustering aims to cluster such fact-sharing sentences in the same communities. Thus, starting with the largest community is important to ensure that the system summary first covers the facts that have higher frequencies and therefore higher weights.
 
 \item {\bf {\it Cluster Lexrank (C-lexrank)}}\\ The second method we used was Lexrank \cite{Erkan&Radev04c} inside each cluster. In other words, for each cluster $\Omega_i$ we made a lexical network of \emph{the sentences in that cluster} ($N_i$) .  Using Lexrank we can find the most central sentences in $N_i$ as salient sentences of $\Omega_i$ to include in the main summary. We simply choose, for each cluster $\Omega_i$, the most salient sentence of $\Omega_i$, and if we have not reached the summary length limit, we do that for the second most salient sentences of each cluster, and so on. The way of ordering clusters is again by decreasing size.
\end{itemize}\vspace{-2mm}

Table \ref{tbl:systemSumm} shows the two system summaries generated with C-RR and C-lexrank methods for P99-1065. The sentences in the table appear as they were extracted automatically from the text files of papers, containing sentence fragments and malformations occurring while doing the automatic segmentation.
\vspace{-2mm}
\subsection{Baseline Methods}

We also conducted experiments with two baseline approaches.
To produce the citation summary we used \emph{Mead}'s \cite{Radev&al.04a} Random Summary  and \emph{Lexrank} \cite{Erkan&Radev04c} on the entire citation summary network as baseline techniques.
Lexrank is proved to work well in multi-document summarization \cite{Erkan&Radev04c}.
It first builds a lexical network, in which nodes are sentences and a weighted edge between two nodes shows the lexical similarity. Once this network is built, Lexrank performs a random walk to find the most central nodes in the graph and reports them as summary sentences.
\vspace{-2mm}
\section{Experimental Setup}

\subsection{Evaluation Method}

Fact-based evaluation systems have been used in several past projects
\cite{jimmy06,Marton-naacl-2006}, especially in the TREC question
answering track. \cite{jimmy06} use stemmed unigram similarity of
responses with nugget descriptions to produce the evaluation results,
whereas \cite{Marton-naacl-2006} uses both human judgments and human
descriptions to evaluate a response.

An ideal summary in our model is one that covers more facts and more
important facts. Our definition for the properties of a ``good''
summary of a paper is one that is relatively short and consists of the
main contributions of that paper. From this viewpoint, there are two
criteria for our evaluation metric.  First, summaries that contain
more important facts are preferred over summaries that cover fewer
relevant facts.  Second, facts should not be equally weighted in this
model, as some of them may show more important contributions of a
paper, while others may not.

To evaluate our system, we use the pyramid evaluation method
\cite{nenkova2004ecs} at sentence level. Each fact in the citation
summary of a paper is a \emph{summarization content unit (SCU)}
\cite{nenkova2004ecs}, and the fact distribution matrix, created by
annotation, provides the information about the importance of each fact
in the citation summary.

The score given by the pyramid method for a summary is a ratio of the
sum of the weights of its facts to the sum of the weights of an
optimal summary. This score ranges from 0 to 1, and high scores show
the summary content contain more heavily weighted facts.  We believe
that if a fact appears in more sentences of the citation summary than
another fact, it is more important, and thus should be assigned a
higher weight. To weight the facts we build a pyramid, and each fact
falls in a tier. Each tier shows the number of sentences a fact
appears in. Thus, the number of tiers in the pyramid is equal to the
citation summary size. If a fact appears in more sentences, it falls
in a higher tier.  So, if the fact $f_i$ appears $|f_i|$ times in the
citation summary it is assigned to the tier $T_{|f_i|}$.

The pyramid score formula that we use is computed as follows. Suppose
the pyramid has $n$ tiers, $T_i$, where tier $T_n$ on top and $T_1$ on
the bottom.  The weight of the facts in tier $T_i$ will be $i$
(i.e. they appeared in $i$ sentences). If $|T_i|$ denotes the number
of facts in tier $T_i$, and $D_i$ is the number of facts in the
\emph{summary} that appear in $T_i$, then the total fact weight for
the summary is $D = \sum_{i=1}^{n} i\times D_i$.  Additionally, the
optimal pyramid score for a summary with $X$ facts, is

$ Max = \sum_{i=j+1}^n i\times |T_i| + j\times (X-\sum_{i=j+1}^n |T_i|)$
where $j= \max_i (\sum_{t=i}^n|T_t|\geq X)$. Subsequently, the pyramid score for a summary is calculated as $ P = \frac{D}{Max} $.
\vspace{-2mm}
\subsection{Results and Discussion}

Based on the described evaluation method we conducted a number of
experiments to evaluate different summaries of a given length.  In
particular, we use a gold standard and a random summary to determine
how good a system summary is.  The gold standard is a summary of a
given length that covers as many highly weighted facts as possible.
To make a gold summary we start picking sentences that cover new and
highly weighted facts, until the summary length limit is reached.  On
the other hand, in the random summary sentences are extracted from the
citation summary in a random manner.  We expect a good system summary
to be closer to the gold than it is to the random one.

Table \ref{tbl:results} shows the value of pyramid score $P$, for the
experiments on the set of 25 papers.  A $P$ score of less than $1$ for
a gold shows that there are more facts than can be covered with a set
of $|S|$ sentences.

This table suggests that C-lexrank has a higher average score, $P$,
for the set of evaluated articles comparing C-RR and Lexrank.

As mentioned earlier in section \ref{sect:extraction}, once the
citation summary network is clustered in the C-RR method, the
sentences from each cluster are chosen in a round robin fashion, which
will not guarantee that a fact-bearing sentence is chosen. 

This is because all sentences, whether they cover any facts or not,
are assigned to some cluster anyway and such sentences might appear as
the first sentence in a cluster. This will sometimes result in a low
$P$ score, for which P05-1012 is a good example.

\begin{table}
 \begin{center}
{\scriptsize
 \begin{tabular}{| c |c| c |  c | c |  c | c | c |}
 \hline
 &  \rotatebox{90}{\mbox{Article}} & \rotatebox{90}{\mbox{Gold}} & \rotatebox{90}{\mbox{Mead's Random}} & \rotatebox{90}{\mbox{Lexrank}} & \rotatebox{90}{\mbox{C-RR}}  & \rotatebox{90}{\mbox{C-lexrank}} \\ \hline
\multirow{5}{*}{\rotatebox{90}{\mbox{DP}}} &  C96-1058  & 1.00 & 0.27 & 0.73 & 0.73 & 0.73 \\
   &  P97-1003  & 1.00 &  0.08 &  0.40 & 0.60 & 0.40 \\
   &  P99-1065  & 0.94 &  0.30 &  0.54 & 0.82 & 0.67 \\
   &  P05-1013  & 1.00 &  0.15 &  0.69 & 0.97 & 0.67 \\
   &  P05-1012  & 0.95 &  0.14 &  0.57 & 0.26 & 0.62 \\
\hline
\multirow{5}{*}{\rotatebox{90}{\mbox{PBMT}}} & N03-1017  & 0.96 & 0.26 & 0.36 & 0.61 & 0.64 \\
     & W03-0301  & 1.00 & 0.60 & 1.00 & 1.00 &  1.00 \\
     & J04-4002  & 1.00 & 0.33 & 0.70 & 0.48 &  0.48 \\
     & N04-1033  & 1.00 & 0.38 & 0.38 & 0.31 &  0.85 \\
     & P05-1033  & 1.00 & 0.37 & 0.77 & 0.77 &  0.85 \\
 \hline
\multirow{5}{*}{\rotatebox{90}{\mbox{Summ}}} &  A00-1043  & 1.00 & 0.66 & 0.95 & 0.71 & 0.95 \\
   &  A00-2024   & 1.00 & 0.26 & 0.86 & 0.73 & 0.60\\
   &  C00-1072   & 1.00 & 0.85 & 0.85 & 0.93 & 0.93\\
   &  W00-0403   & 1.00 & 0.55 & 0.81 & 0.41 & 0.70\\
   &  W03-0510  & 1.00  & 0.58 & 1.00 & 0.83 & 0.83\\
\hline
\multirow{5}{*}{\rotatebox{90}{\mbox{QA}}} &  A00-1023 & 1.00 & 0.57 & 0.86 & 0.86 & 0.86 \\
   &  W00-0603  & 1.00 & 0.33 & 0.53 & 0.53 & 0.60\\
   &  P02-1006  & 1.00 & 0.49 & 0.92 & 0.49 & 0.87\\
   &  D03-1017  & 1.00 & 0.00 & 0.53 & 0.26 & 0.85\\
   &  P03-1001  & 1.00 & 0.12 & 0.29 & 0.59 & 0.59\\
 \hline
\multirow{5}{*}{\rotatebox{90}{\mbox{TE}}} &    D04-9907 & 1.00 & 0.53 & 0.88 & 0.65 & 0.94\\
   &  H05-1047 & 1.00 & 0.83 & 0.66 & 0.83 & 1.00 \\
   &    H05-1079 & 1.00 & 0.67 & 0.78 & 0.89 & 0.56\\
   &    W05-1203 & 1.00 & 0.50 & 0.71 & 1.00 & 0.71\\
   &    P05-1014 & 1.00 & 0.44 & 1.00 & 0.89 & 0.78\\
\hline
 & {\bf  Mean} & {\bf 0.99} &  {\bf 0.41} & {\bf 0.71} & {\bf 0.69} & {\bf  0.75 }\\
\hline
 \end{tabular}}
 \end{center}
 \caption{Evaluation Results ($|S|=5$)}\label{tbl:results}
\vspace{-2mm}
 \end{table}

\vspace{-3mm}
\section{Conclusion and Future Work}

In this work we use the citation summaries to understand the main
contributions of articles. The citation summary size, in our
experiments, ranges from a few sentences to a few hundred, of which we
pick merely a few ($5$ in our experiments) most important ones.

As a method of summarizing a scientific paper, we propose a clustering
approach where communities in the citation summary's lexical network
are formed and sentences are extracted from separate clusters. Our
experiments show how our clustering method outperforms one of the
current state-of-art multi-document summarizing algorithms, Lexrank,
on this particular problem.

A future improvement will be to use a reordering approach like Maximal
Marginal Relevance (MMR) \cite{Carbonell&Goldstein98} to re-rank
clustered documents within each cluster in order to reduce the
redundancy in a final summary. Another possible approach is to assume
the set of sentences in the citation summary as sentences talking
about the same event, yet generated in different sources. Then one can
apply the method inspired by \cite{Barzilay&al.99} to identify common
phrases across sentences and use language generation to form a more
coherent summary.  The ultimate goal, however, is to produce a topic
summarizer system in which the query is a scientific topic and the
output is a summary of all previous works in that topic, preferably
sorted to preserve chronology and topicality. 

\vspace{-3mm}
\section{Acknowledgments}

The authors would like to thank Bonnie Dorr, Jimmy Lin, Saif
Mohammad, Judith L. Klavans, Ben Shneiderman, and Aleks Aris from UMD, Bryan Gibson, Joshua Gerrish,
Pradeep Muthukrishnan, 
Arzucan \"Ozg\"ur, Ahmed Hassan, and Thuy Vu from
University of Michigan for annotations.

This paper is based upon work supported by the National Science
Foundation grant "iOPENER: A Flexible Framework to Support Rapid
Learning in Unfamiliar Research Domains", jointly awarded to U. of
Michigan and U. of Maryland as IIS 0705832. Any opinions, findings,
and conclusions or recommendations expressed in this paper are those
of the authors and do not necessarily reflect the views of the
National Science Foundation.

\vspace{-5mm}

\bibliographystyle{coling}
\bibliography{ref}

\end{document}